\documentclass[a4paper,12pt]{article}
\usepackage{cite}
\usepackage{amssymb}
\usepackage{multicol}
\usepackage{amsmath}
\usepackage{pgfplots}
\usepackage{mathrsfs} % for \mathscr
\usepackage{xfp} % higher precision (16 digits?)
\usepackage[outline]{contour} % glow around text

\usepackage{epsfig}
\usepackage{graphicx}
\usepackage{slashed}
\usepackage{xcolor,color}
\usepackage{ulem}
\usepackage{subfigure}
\usepackage[toc,page]{appendix}
\usepackage[makeroom]{cancel}
\usepackage{tikz}
\usetikzlibrary{shapes.geometric,positioning,decorations.pathreplacing,decorations.pathmorphing,patterns,decorations.markings}

\usepackage{scalerel}
\usepackage[usestackEOL]{stackengine}

\def\dashint{\,\ThisStyle{\ensurestackMath{%
            \stackinset{c}{.2\LMpt}{c}{.5\LMpt}{\SavedStyle-}{\SavedStyle\phantom{\int}}}%
        \setbox0=\hbox{$\SavedStyle\int\,$}\kern-\wd0}\int}

\makeatletter
\renewcommand{\theequation}{\thesection.\arabic{equation}}
\@addtoreset{equation}{section}
\makeatother
\def\be{\begin{equation}}
\def\ee{\end{equation}}
\def\bea{\begin{eqnarray}}
\def\eea{\end{eqnarray}}
\def\bgn{\begin{align}}
\def\egn{\end{align}}
\def\nn{\nonumber \\}
\def\({\left(}
\def\){\right)}
\def\<{\left<}
\def\>{\right>}

\def\({\left(}
\def\){\right)}
\def\<{\left<}
\def\>{\right>}
\def\!{\right|}
\def\|{\left|}

\def\[{\left[}
\def\]{\right]}

\def\+{\bar}

\def\W{{\cal{W}}}

\def\C{{\cal{C}}}

\def\i{{\bar{i}}}

\usepackage{hyperref}
\hypersetup{colorlinks,%
  citecolor=blue,%
  linkcolor=blue,%
  pdftex}

\begin{document}

\begin{titlepage}
\vskip1cm
\begin{flushright}
% UOSTP {\tt 1512001}
\end{flushright}
\vskip0.25cm
\centerline{
\bf \large 
Resolving Black Hole Singularities in Jackiw-Teitelboim Gravity
} 
\vskip0.8cm \centerline{ \textsc{
 Dongsu Bak,$^{ \negthinspace  a}$  Chanju Kim,$^{ \negthinspace b}$ Sang-Heon Yi,$^{\negthinspace a}$} }
\vspace{0.8cm} 
\centerline{\sl  a) Physics Department \& Natural Science Research Institute}
\centerline{\sl University of Seoul, Seoul 02504 \rm KOREA}
 \vskip0.2cm
 \centerline{\sl b) Department of Physics, Ewha Womans University,
  Seoul 03760 \rm KOREA}
%   \vskip0.2cm
% \centerline{\sl c) Center for Quantum Spacetime \&  Physics Department}
%  \centerline{\sl Sogang University,  Seoul 04107 \rm KOREA}
\vskip0.4cm

 \centerline{
\tt{(\small dsbak@uos.ac.kr,\,cjkim@ewha.ac.kr,\,shyi704@uos.ac.kr})
} 
  \vspace{1.5cm}
%\centerline{\today}
%\vspace{1.75cm}
\centerline{ABSTRACT} \vspace{0.65cm} 
{
\noindent
In Jackiw-Teitelboim  gravity, the naive Schwarzian quantum mechanics leads to a continuous bulk spectrum, in apparent contradiction with the finite entropy of the black hole, which requires a discrete spectrum with level spacing of order $e^{-S_0}$. It was recently shown that restoring spectral discreteness  with random statistics requires the introduction of a left confining potential that becomes relevant when the renormalized wormhole length reaches order $e^{S_0}$. In this work, we show how the known perturbative results of JT gravity are  recovered within this modified framework. More importantly, we demonstrate that this modification has a direct dynamical consequence: it resolves the black-hole singularity. The confining potential generates a repulsive force at exponentially large wormhole length, preventing the indefinite growth that would otherwise lead to a singularity.
We explain in detail how this turnaround arises and explore its implications for late-time bulk gravitational dynamics, the disappearance of horizons, and possible observational consequences.

}

%\vspace{0.75cm}
%\centerline{(\today)}
\end{titlepage}
%%%%%%%%%%%%%%%%%%%%%%
%\maketitle

%%%%%%%%%%%%%%%%%

%%%%%%%%%%%%%%%%%%%%%%%%%
\section{Introduction}\label{sec1}
%\section{Outline}\label{sec0}
%%%%%%%%%%%%%%%%%%%%%%%%%
%%%%%%%%%%%%%%%%%%%%%%%%%
%Black holes, once regarded as purely academic objects, have been observed astrophysically. 
 In Jackiw-Teitelboim (JT) gravity~\cite{Jackiw:1984je,Teitelboim:1983ux,Almheiri:2014cka}, the vacuum solution describes a Lorentzian two-sided black hole, which corresponds to the disk geometry on the Euclidean side~\cite{Saad:2019lba}. The Lorentzian spacetime contains a wormhole connecting the left and right asymptotic boundaries, and its renormalized length (or, in higher dimensions, the renormalized volume) provides a gauge-invariant geometric observable~\cite{Susskind:2018pmk,Susskind:2014moa,Susskind:2014rva,Brown:2018bms,Susskind:2019ddc,Susskind:2020gnl}. Upon quantization, this degree of freedom reduces at leading order to Schwarzian quantum mechanics~\cite{Maldacena:2016upp,Harlow:2018tqv,Bagrets:2016cdf}, whose wave function depends on 
the variable $\chi=- \ell_{ren}/2$ with 
$\ell_{ren}$ denoting the renormalized wormhole length.

 The black hole carries a finite number of degrees of freedom, and therefore its energy spectrum is expected to be discrete, with a typical level spacing of order $e^{-S_0}$. However, in the Schwarzian description the dynamical variable $\chi$ ranges over the entire real line, 
and the Liouville  potential $e^{2\chi}$ is not confining.  As a result, the corresponding quantum-mechanical spectrum is continuous~\cite{Stanford:2017thb}. This is in contradiction with the finiteness of the black hole entropy, which demands a discrete spectrum.

 In our previous work~\cite{Bak:2025eul}, we resolved this issue by introducing a left confining potential that becomes $O(1)$ only when $\chi = -O(e^{S_0})$. Consequently, the potential effectively disappears in the limit $S_0$ goes to infinity, which corresponds to the continuum limit familiar from semiclassical gravity.  At leading order, the left confining potential $W(X)$, with the rescaled variable $X=-2\chi e^{-S_0}$, is determined explicitly through a semiclassical analysis that reproduces the well-known disk density of states~\cite{Mertens:2022irh,Turiaci:2024cad}. Higher-order corrections are incorporated by introducing random components into the potential, capturing the subleading effects beyond the semiclassical approximation.

 This modification leads to a dramatic change in the gravitational dynamics once the wormhole length reaches $O(e^{S_0})$, an exponentially large scale. The wormhole length can be identified with the so-called Krylov spread complexity~\cite{Balasubramanian:2019wgd,Balasubramanian:2022tpr,Balasubramanian:2022dnj,Erdmenger:2023wjg,Rabinovici:2023yex,Balasubramanian:2024lqk}.
For a finite-temperature quantum-mechanical system with a discrete spectrum and random level statistics, it is well known that the Krylov spread complexity initially grows     exponentially  in time, accompanied by a universal linear growth,  reaches a maximum, and then decreases before eventually saturating at a plateau. In this late-time regime, the system becomes effectively maximally random and highly complex   (see~\cite{Nandy:2024evd,Baiguera:2025dkc,Rabinovici:2025otw} for reviews of Krylov spread complexity).

In this note, we examine the implications of introducing the left confining potential. A particularly striking consequence is the resolution of the black hole singularity. We explain in detail how this resolution arises and discuss its physical implications.
In essence, the singularity is resolved because the left confining potential generates a repulsive force that counteracts the otherwise purely attractive gravitational dynamics. Since this confining potential originates from the discreteness of the energy spectrum, it follows that spectral discreteness itself is ultimately responsible for the singularity resolution. In this sense, the mechanism is intrinsically nonclassical.
In the absence of the left confining potential, the linear growth of the wormhole length up to times
$t \sim O(e^{S_0})$ would inevitably lead to a singularity. Up to this regime, gravity remains purely attractive and exhibits the familiar features of chaotic dynamics, thermalization, and scrambling on $O(1)$ timescales, reflecting its thermodynamic character. By contrast, the regime near the maximum of the complexity—the top, the subsequent downward slope, and the eventual  plateau—corresponds to  exponentially long timescales that are usually neglected. It is precisely in this regime that the effects of spectral discreteness become dominant and qualitatively modify the bulk dynamics.

Section~\ref{sec2} reviews %Jackiw–Teitelboim (
JT gravity and explains how the singularity arises in the standard formulation. We also briefly summarize the Saad–Shenker–Stanford (SSS) duality~\cite{Saad:2019lba}. In Section~\ref{sec3}, we introduce the left confining potential on the JT side and explain how the perturbative results are recovered within the modified JT framework. Section~\ref{sec4} demonstrates how the singularity is resolved, while the corresponding bulk interpretation is presented in Section~\ref{sec5}. The final section is devoted to conclusions and discussions. In the Appendix, we show how the perturbative genus expansion emerges in our framework and demonstrate explicitly  that its first few terms satisfy nontrivial consistency conditions.

\section{Bulk singularity from boundary dynamics}\label{sec2}
%%%%%%%%%%%%%%%%%%%%%%%%%
%%%%%%%%%%%%%%%%%%%%%%%%%
%Black holes, once regarded as purely academic objects, have been observed astrophysically. 
We   begin this section  with  the  bulk JT gravity action, 
\begin{equation} \label{}
I = \frac{1}{16\pi G}\int_{M} d^{2}x\sqrt{-g}~\phi (R+2) + \frac{1}{8\pi G}\int_{\partial M}dt \sqrt{-\gamma_{tt}}~  \phi (K-1) \,,
\end{equation}
where $\phi$ denotes a dilaton field, $t$ represents the physical boundary time, 
%at the cutoff trajectories,  
and $\gamma_{tt}$ and $K$ are the induced metric and the extrinsic curvature on the boundary $\partial M$, respectively.  In the two-sided Lorentzian setup, the boundary term contains contributions from both the left and right asymptotic boundaries.  Varying the action with respect to the dilaton and the metric yields the bulk equations of motion:
\begin{align}    \label{}
0 &= R+2\,,  \\
0 &=  \nabla_{a}\nabla_{b}\phi - g_{ab} \nabla^{2}\phi + g_{ab}\phi \,. 
\end{align}
The first equation  dictates the bulk spacetime to be  locally AdS$_{2}$.  In the Euclidean framework, we can also  include topological terms, 
\begin{equation} \label{}
I^{\rm top}_{E} = - \frac{S_{0}}{2\pi} \frac{1}{8\pi G}\bigg[\frac{1}{2}\int_{M_{E}}\sqrt{g}R + \int_{\partial M_{E}}\sqrt{\gamma}K\bigg]\,, 
\end{equation}
which can be used for the natural topological expansion by powers of $e^{-S_0}$.
Below we shall set $8\pi G=1$ for the simplicity of our presentation.

Now, we briefly review the boundary dynamics of the  AdS$_2$ geometry. 
We work with the global $AdS_{2}$ metric 
\be \label{global}
ds^2= \frac{-d\tau^2+d\mu^2}{\cos^2\mu}\,,
\ee
with $\mu \in [-\pi/2, \pi/2]$ where $\mu=\pm\pi/2$ correspond to the left and right spatial infinities, respectively. 
The cutoff trajectory is defined by fixing the value of the dilaton as
\be\label{cutoff}
\phi(\tau,\mu)|_{c} =\frac{\bar{\phi}}{\epsilon}\,,
\ee
following Refs.~\cite{Almheiri:2014cka,Maldacena:2016upp,Jensen:2016pah,Engelsoy:2016xyb,Mertens:2017mtv,Kitaev:2017awl}.
The boundary physical time $t$ is determined by requiring that the induced metric on the cutoff trajectory takes the form 
\be\label{repara}
ds^2|_{c}= -\frac{dt^2}{\epsilon^2}\,.
\ee
%{\color{yellow}We work with the global $AdS_{2}$ metric 
%\be \label{global}
%ds^2= \frac{-d\tau^2+d\mu^2}{\cos^2\mu}\,,
%\ee
%with $\mu \in [-\pi/2, \pi/2]$ where $\mu=\pm\pi/2$ correspond to the left and right spatial infinities, respectively. }
The left and right cutoff trajectories are determined by embedding functions $\tau=\tau(t)$ and $\mu=\mu_c(t)$. 
Imposing the induced metric condition yields
\be
\cos \mu_c(t)= \epsilon \, \dot{\tau}\,,
\ee
where the dot denotes differentiation with respect to the physical boundary time $t$. 
Now suppose the dilaton admits the near-boundary expansion
\be\label{dilatona}
\phi =\bar\phi \left(\frac{G(\tau)}{\cos \mu} + O(\cos \mu)\right)\,.
\ee
Evaluating this expression on the cutoff trajectories and using condition~\eqref{cutoff}, we arrive at the simple  boundary relation
\be \label{tauprime} 
\dot\tau= G(\tau)\,.
\ee
Thus, the boundary time reparametrization is determined directly by the leading near-boundary behavior of the dilaton. %The relations form the starting point for deriving the effective Schwarzian boundary dynamics.
This relation encodes how the bulk dilaton profile fixes the boundary dynamics.

With this preparation, the left and right boundary Lagrangians take the Schwarzian form
\be
L_{r/l}=-{\cal C}\,  {\rm Sch}\left(\tan \frac{\tau_{r/l}(t)}{2}, t\right) \,,
\ee 
 where ${\cal C}=\bar\phi$ and  the Schwarzian derivative is defined by
\be
{\rm Sch}(f(x), x) =-\frac{1}{2}(f''/f')^2 +(f''/f')'\,.
\ee
It is convenient to rewrite the Schwarzian action in a first-order form,
\be
 L_{r/l}=
\frac{\cal C}{2} ({\dot\chi}_{r/l})^2 - \frac{1}{2{\cal C}} e^{2\chi_{r/l}}+p_{r/l} \left(\dot\tau_{r/l}-\frac{1}{\cal C}
e^{\chi_{r/l}}\right)\,,
\ee
where $p_{r/l}$ plays a role of Lagrange multiplier.
After fully fixing the large gauge redundancy and quantizing the constrained system (see, for example, \cite{Bak:2023zkk} for details), %the total boundary Lagrangian reduces to
the left and right sectors become identical and the total boundary Lagrangian reduces to
\be\label{total}
L_{total}= L_{r}+L_{l}=%\frac
%{\cal C} ({\chi'}_{r/l})^2 - \frac{1}{{\cal C}} e^{2\chi_{r/l}}
{\cal C} ({\dot\chi})^2 - \frac{1}{{\cal C}} e^{2\chi}
\ee
with $\chi=\chi_l=\chi_r$ and ${p_\chi}/2=p_{\chi_r}=p_{\chi_l}$.  
%These are basically from the quantization of the constraint system and  the corresponding  gauge fixing condition, whose full details are presented for instance in \cite{Bak:2023zkk}.
Also note that $H_{total}/2=H_l=H_r=H$ where $H_{total}$ is the Hamiltonian derived from the above Lagrangian (\ref{total}). In particular,  the Hamiltonian $H$ takes the form
\be \label{SchHam}
{2\C} H=\frac{1}{4}p_{\chi}^2+e^{2\chi}=p_q^2+e^q
\ee
with $q=2\chi$.
The variable $\chi$ is defined with the relation $e^\chi={\cal C}\dot\tau$ and one can show that the renormalized wormhole length $\ell_{ren}$ is given by $\ell_{ren}=-2\chi$ \cite{Bak:2023zkk}.  
%From now on, we shall set ${\cal C}=1$ for the simplicity of our presentation.

With the relation (\ref{tauprime}), let us compare the resulting equation of motion.
First note that $\dot\chi= \dot\tau {G'}/G$ with  a prime here denoting a $\tau$ derivative 
and, if we use (\ref{tauprime}) again, one has
$\dot\chi ={G'}$. With one more $t$ derivative, we are led to 
\be\label{orieom}
 \C {\ddot\chi} = \C {G''} \dot\tau=  \C G {G''}= -\frac{e^{2 \chi}}{\C}\,,
\ee
where we use the equation of motion for the last equality. Noting $e^\chi=\C G(\tau)$, one finds 
$ {G''}= -G$. This is precisely the equation satisfied by the vacuum black hole solution,
%This is indeed consistent with the vacuum black hole solution with 
$G= \cos\tau$, demonstrating the consistency between the Schwarzian boundary dynamics and the bulk dilaton solution.

Let us analyze this vacuum solution with $\dot\tau= \cos\tau$. Solving this equation gives
\be
\dot\tau=\cos\tau=1/\cosh t\,,
\ee
where we have chosen the integration constant such that $\tau(0)=0$.
As $t$ goes to infinity, the velocity $\dot\tau  \rightarrow 0$ and $\tau \rightarrow \pi/2$. With
$\C \dot\tau=e^\chi$, $\ell_{ren}=-2\chi \sim 2t -2\log 2{\cal C}\rightarrow \infty$. This statement refers to the fixed branch selected by $\tau(0)=0$. Other branches of the global AdS$_2$ cover are obtained by constant shifts of $\tau$ and have shifted endpoints, but the branch label is not a dynamical quantity that grows during the time evolution. Thus the unmodified vacuum solution is bounded on each fixed branch; for the branch used here, the endpoint is $\pi/2$. Nevertheless, the wormhole length and the corresponding complexity grow indefinitely. This indefinite growth is closely tied to the appearance of a singularity in the geometry.
%: {\color{red}along the boundary trajectory, $\tau$ approaches the endpoint $\pi/2$ while the renormalized wormhole length diverges as $t\to\infty$.}
%This indefinite growth is closely tied to the appearance of a singularity in the geometry, since $\tau$ is bounded by $\pi/2$ along the boundary trajectory as time $t$ goes to infinity. 

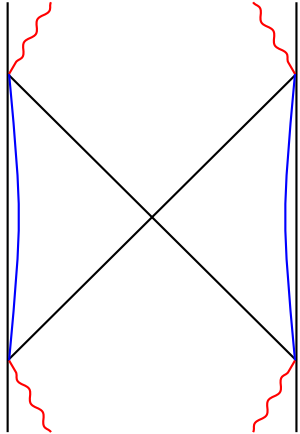
\begin{figure}  
\vskip0.3cm 
\begin{center}
%
%
%
%%%%%%%%%%%%%%%%%%%%%%%%%%%%%%%%%%%%%%%%%%%%%%%%
\begin{tikzpicture}[scale=1.9]
%\draw[thick,blue %,-<
%](-2,0)--++(0,2);
%\draw[thick,blue] (-2,2)%node[left]{\large$t_l$}
%--++(0,2);
%\draw%[thick]
%(-3,0)--++(0,3);
%\draw%[thick]
%(-1.3,0)--++(0,3);
\draw[thick]
(1.7,0)--(1.7,3);
\draw[thick]
(3.7,0)--(3.7,3);
\draw[thick] (3.7,0.5) -- (1.7,2.5) ;
\draw[thick] (1.7,0.5) -- (3.7,2.5) ;

\draw[thick,blue] 
(1.71,0.5) .. controls (1.8,1.5) .. (1.71,2.5);
\draw[thick,blue] 
(3.69,0.5) .. controls (3.6,1.5) .. (3.69,2.5);

\draw[thick,red,decorate, decoration={snake,amplitude=.4mm}] (2.0,3.0) -- (1.71,2.5) ;
\draw[thick,red,decorate, decoration={snake,amplitude=.4mm}] (3.4,3.0) -- (3.69,2.5) ;
\draw[thick,red,decorate, decoration={snake,amplitude=.4mm}] (2.0,0) -- (1.71,0.5) ;
\draw[thick,red,decorate, decoration={snake,amplitude=.4mm}] (3.4,0) -- (3.69,0.5) ;
\end{tikzpicture}
%%%%%%%%%%%%%%%%%%%%%%%%%%%%%%%%%%%%%%%%%%%%%%%%%%%%%%%%%%

\end{center} 
\vskip-0.3cm 
\caption{Penrose diagram of bulk AdS$_{2}$ spacetime with horizons and boundary cutoff trajectories.} \label{fig1}
\end{figure}

%The boundary physical system terminates at 
The crucial feature of this solution is that the boundary time evolution terminates at the limiting value
$\tau=\pi/2$. Although this limit is reached only asymptotically in $t$, the global coordinate 
$\tau$ itself is bounded. The boundary trajectory therefore accumulates at a finite value of global time $\tau$ while the renormalized geodesic length diverges. In the bulk picture, the future left/right horizons are located at $\tau=\mp \mu$. The horizons intersect the boundary precisely at
$\tau=\pi/2$. Consequently, as 
$t\rightarrow \infty$, the boundary trajectory approaches the endpoint of the future horizon.
Because the evolution in 
$\tau$ cannot extend beyond 
$\pi/2$, the bulk spacetime develops a causal pathology behind the horizon. This manifests as the familiar black-hole singularity in the interior region. This well-known bulk black hole geometry is depicted in Figure \ref{fig1}. 

%Consequently, the corresponding bulk spacetime inevitably develops a pathological behavior due to causality, resulting in a singularity inside the black-hole horizon.  In this case, the left/right future horizon given by $\tau=\mp \mu$ ends at $\tau=\pi/2$ and, hence, the bulk trajectory remains outside the black hole horizon. This well-known bulk black hole geometry is depicted in Figure \ref{fig1}. 

We now briefly review the connection between the boundary 
description of JT gravity discussed above and the 
SSS duality~\cite{Saad:2019lba}. The SSS duality can be viewed as providing a nonperturbative completion of 
JT gravity in terms of an appropriate random matrix model.

When the leading JT bulk dynamics is governed by the %Liouville-type 
Schwarzian Hamiltonian in~\eqref{SchHam}, the corresponding eigenstates are scattering states described by modified Bessel functions, and the spectrum is continuous. This reflects the absence of a discrete density of states in the Lorentzian Schwarzian %Liouville 
description.
According to the %Saad–Shenker–Stanford ()
SSS duality, however, this apparent continuum should be understood as emerging from an underlying discrete spectrum with typical level spacing of order $e^{-S_{0}}$.  In the Euclidean framework, the JT disk density of states, $\rho_{JT}(E) = \frac{e^{S_{0}}}{4\pi^2}\sinh (2\pi\sqrt{E})$ was derived in~\cite{Stanford:2017thb,Saad:2019lba} and subsequently used as input for computing higher-genus corrections in the double-scaling limit of the associated random matrix model.
Nevertheless, within the Liouville potential alone, genuine spectral discreteness cannot be realized, as noted above. This apparent clash motivates the introduction of an additional left confining potential, as proposed in~\cite{Bak:2025eul}, which restores discreteness while incorporating the random nature of the spectrum. In the following section, we review this left confining potential and discuss its immediate dynamical consequences.

%%%%%%%%%%%%%%%%%%%%%%%%%
%\section{Introduction}\label{sec0}
\section{Confining potential and perturbative framework}\label{sec3}
%%%%%%%%%%%%%%%%%%%%%%%%%
%%%%%%%%%%%%%%%%%%%%%%%%%
%In this section, we review the left confining potential, formulate the corresponding modified %version of JT theory that incorporates the additional potential, and examine its physical implications. 
In this section, we review the left confining potential and formulate a modified version of JT gravity that incorporates this additional potential.

On the matrix model side (see for review~\cite{Mehta2004,Eynard:2015aea,Cotler:2016fpe}), we consider an $N\times N$ Hermitian matrix $M$ with eigenvalues $\lambda_i \,(i=1,2,\cdots, N)$.  Defining $\widehat{Z}  = \text{tr}\, e^{- \beta M}=\sum_i e^{-\beta \lambda_i}$, the averaged matrix-model partition function is obtained by averaging over the eigenvalue ensemble,
\be
\langle \widehat{Z}\rangle_{\vec{\lambda}}\equiv \int d\vec\lambda\, {\cal P}(\vec\lambda) \, \widehat{Z} (\vec\lambda)\,,
\ee
where ${\cal P}(\vec\lambda)$ is the appropriately normalized JT ensemble weight satisfying $\int d\vec\lambda\, {\cal P}(\vec\lambda) =1$. Throughout this work, we take the double-scaling limit in which $N \to \infty$ while the level-spacing parameter $e^{-S_0}$ is held fixed and small. In the limit, the averaged matrix-model partition function admits the genus expansion
as\footnote{This can be generalized to the connected correlation function  of  $\widehat{Z}(\beta_1), \widehat{Z}(\beta_2), \cdots,  \widehat{Z}(\beta_n) $,  which has the expansion
\be
\langle \widehat{Z}(\beta_1)\widehat{Z}(\beta_2)\cdots  \widehat{Z}(\beta_n) \rangle_{\vec{\lambda},conn} \cong  \sum^\infty_{g=0} e^{-(2g-2+n)S_0} Z_{g,n}(\beta_1,\beta_2,\cdots,\beta_n)\,.
\ee
 }
\be\label{genusn}
\langle \widehat{Z}\rangle_{\vec{\lambda}} \cong e^{S_0} Z_{0,1} +\sum^\infty_{g=1} e^{-(2g-1)S_0} Z_{g,1} \,, 
\ee 
where $\cong$ indicates equality up to nonperturbative contributions, which are not yet fully understood \cite{Mertens:2022irh}. The leading semiclassical contribution $e^{S_0} Z_{0,1}$ can be computed from the JT disk density of states $\rho_{JT}$ via a simple Laplace transform, and all remaining higher-order terms can be determined by the so-called topological recursion relations~\cite{Eynard:2004mh,Mirzakhani:2006fta,Eynard:2007kz,Eynard:2007fi}. 
Under the SSS duality~\cite{Saad:2019lba}, the matrix-model expansion described above is mapped term by term to the JT gravity path integral. In this correspondence, $Z_{0,1}$ reproduces the disk amplitude, while $Z_{g,1}$ captures the contributions from genus-$g$ geometries, which we do not review further here.

\begin{figure}[htbp] 
\begin{center}
\begin{tikzpicture}[scale=1.5]
\begin{axis}[
        axis lines =  center ,
        axis y line=none,
        axis x line=none,
        xlabel = $x$,
        ylabel = {$V(q)$},
        xmin=-20, xmax=5,
        ymin=-0.3, ymax=1,
        xtick=\empty,
        ytick=\empty,
        xlabel style={below right},
        ylabel style={left above},
        clip=false,
        y=0.8cm
     ]
       
    % Function curve (tanh-like, shifted up)
    \addplot[blue, smooth, domain=-15:4.6, samples=200] {250*exp(x-9.8)+0.015};
     \node[font=\tiny] at (axis cs:5.5,1.4)  {$e^{q}$};
    %\addplot[thick, blue, smooth, domain=-15:-12.5, samples=200] {-5*x+0.01};
     \addplot[blue, smooth, domain=-20:-14.98, samples=200] {-2.4*log10(log10(x+22.5))- 0.13};
   % \addplot[blue,smooth,thick] coordinates {(-20,1.3) (-18,0.8)   (-15,0.4)   (-12.8,0.1)     (-12.3,0.05)     (-12.1,0.03)      (-11.98,0.014) };
    \draw[->] (axis cs:2.5,-0.2) -- (axis cs:2.5,1.6)  ;
    \node[font=\tiny] at (axis cs:2.1,1.68)  {$V$};
     \draw[->] (axis cs:-20,0) -- (axis cs:5,0)  ;
     \node[font=\tiny] at (axis cs:5.3,-0.15)  {$q$};

    % Labels
  %  \node[below left] at (axis cs:2.5,0) {$O$};
    
    % Point Q
    \draw[dotted,thick] (axis cs:-18.5,0) -- (axis cs:-18.5,0.4);
   
    \node[font=\tiny,below] at (axis cs:-17,0) {$q\sim {\cal O}(-e^{S_0}%\text{$S_{0}$}
) $};
    
    % Additional label
    %\node[right] at (axis cs:4,1.7) {$b^2$};
 %  
\end{axis}
\end{tikzpicture}
\end{center}
\caption {A schematic form of the potential $V(q)=e^q+W(q)$, where the left confining potential $W(q)$  becomes $O(1)$ only when  $q$ becomes of $-O(e^{S_0})$. With the left confining potential, the spectrum becomes discrete. 
}\label{fig2}
\end{figure}
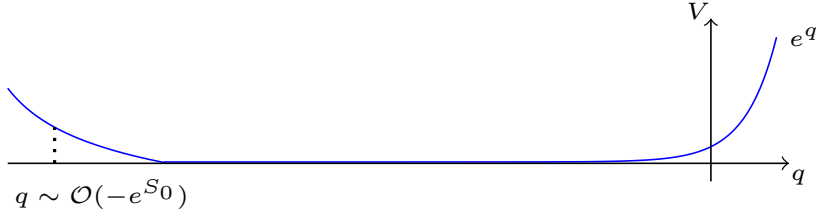

As discussed in~\cite{Bak:2025eul}, the Schwarzian Hamiltonian in~(\ref{SchHam}) gives rise to a continuous spectrum because the Liouville potential $e^{q}$ is not confining as $q \to -\infty$. (Throughout this section and the Appendix, we set $2\mathcal{C}=1$ for notational simplicity.) To obtain a discrete spectrum, we supplement the system with an additional left confining potential
\be
W(X)= W_0(X) + \sum^\infty_{n=1} v_n(X)  W_n(X) \,e^{-n S_0}\,, \quad \quad X=-q\, e^{-S_0}
\ee
(see Figure \ref{fig2}). Here $W_0(X)$ denotes the leading-order confining potential. 
It becomes effective only in the regime $q \sim -O(e^{S_0})$, {\it i.e.}  exponentially far to the left in the original Schwarzian coordinate. 
The functions $v_n(X)$ represent random potentials, introduced in order to reproduce the random nature of the level spacing\footnote{The product $v_n(X)W_n(X)$ may equivalently be replaced by a single random function $w_n(X)$. The separate introduction of the weight function $W_n(X)$ is merely a matter of convenience.}.
With this additional left confining potential, the spectrum becomes discrete once $e^{-S_0}$ is kept finite. 
The corresponding Hamiltonian takes the form
\be
H= e^{-2 S_0} p^2_X +e^{-X e^{S_0}} + W(X)\,,
\ee
with $p_X= -i \partial_X$. We now define a change of variables from the matrix eigenvalues $\vec{\lambda}$ to the collective variables $\{X,\vec{v}(X)\}$ such that
\be
\langle \widehat{Z}(\beta_1)\cdots \widehat{Z}(\beta_n) \rangle_{\vec{\lambda}}=
\langle Z_H(\beta_1)\cdots Z_H(\beta_n) \rangle_{\vec{v}}\equiv \int {\cal D}\vec{v}\, e^{-{\cal U}(\vec{v})} Z_H(\beta_1) \cdots Z_H(\beta_n) 
\ee
where $Z_H (\beta) = {\rm tr}\, e^{-\beta H}$ is the quantum-mechanical partition function associated with the Hamiltonian $H$. 

The statistical properties of the random potentials are encoded in the weight functional. 
%{\color{red}The one point function $\langle v_m(X)\rangle_{\vec{v}}=0$ %is set to be zero 
%as  we choose $W_0$ as a saddle point solution.}
%At the level of 
For two-point correlations, we take ${\cal U}_2 
=\frac{1}{2}\sum_{mn} \int dXd Y v_m(X) C^{mn}(X,Y) v_n(Y)$, which implements correlations,
\be 
  \langle v_m(X) v_n(Y)\rangle_{\vec{v}}=%\delta_{mn} 
C_{mn}(X,Y)\,,
\ee
%with $\langle v_m(X)\rangle_{\vec{v}}=0$ 
where  $\langle v_m(X)\rangle_{\vec{v}}=0$ and $C_{mn}(X,Y)$ is an inverse kernel of $C^{mn}(X,Y)$.
%\sout{We require the correlation function $C(X-Y)$ to be translation invariant.}  
In general, specifying only the two-point functions is not sufficient: 
additional input is required to determine higher-point correlation functions of the random potentials. These higher cumulants encode further statistical data of the ensemble and must be provided in order to fully define the theory beyond leading Gaussian order.

With $\hbar\equiv e^{-S_0}$, the term $e^{-X/\hbar}$ is intrinsically nonperturbative in $\hbar$. As demonstrated in the Appendix, this term generates an infinitely steep potential barrier in the limit $\hbar \to 0$, effectively excluding the region $X<0$ from the physical configuration space. 
In other words, its only role is to enforce a boundary condition at $X=0$.
Importantly, for $X>0$ the dynamics are unaffected at any finite order in the perturbative expansion in $\hbar$. 
%Since $e^{-X/\hbar}$ admits no power-series expansion in $\hbar$, it does not contribute to the perturbative $\hbar$-expansion of observables. 
Consequently, when focusing exclusively on the perturbative expansion in powers of $\hbar$, it is consistent to neglect this nonperturbative contribution and restrict the configuration space to the half-line $X\in [0,\infty)$.

The function  
 $C_{11}(X,Y)$ %$C(X-Y)$ 
can be reconstructed from the connected two-boundary partition function $Z_{0,2}(\beta_1,\beta_2)$; the details of this derivation are provided in the Appendix. The resulting expressions are
\be
%\color{red}
C_{11}(X,Y)
=\delta (X-Y)\,, \quad \quad W_1 (X)=\sqrt{2W'_0(X)} \,,
\label{cw1}
\ee
where  the particular choice of the weight function $W_1(X)$ is made only for convenience. %fix the normalization of $C(X)$ to be unity. 
 Proceeding to the next order, the connected three-boundary amplitude
$Z_{0,3}(\beta_1,\beta_2,\beta_3)$ determines the three-point correlation function of the random potential. One finds
\be %\color{red}
 \langle v_1(X) v_1(Y) v_1(Z)\rangle_{\vec{v}}=\hbar %e^{-S_0} 
\, \frac{ 2^2\,W''_0(X)}{(W_1(X))^3} \delta(X-Y)\delta(Y-Z)\,,
%\,Q_\epsilon(X)Q_\epsilon(Y) Q_\epsilon(Z)
\label{threep}
\ee
whose details %derivation 
are again presented in the Appendix.
%\be
% \langle v_1(X) v_1(Y) v_1(Z)\rangle_{\vec{v}}=-e^{-S_0}\,Q_\epsilon(X)Q_\epsilon(Y) %Q_\epsilon(Z)
%\label{threep}
%\ee
%where
%\be
%Q_\epsilon(X)=\epsilon W_1(X) W_0^{\epsilon-1}(X) = W_1(X) \frac{d W^\epsilon_0}%{dW_0}\,.
%\ee
%\sout{Here $\epsilon$ is the regularization parameter and the limit $\epsilon\rightarrow 0$ should be  taken at the end of the computation.}  
In principle, all higher-point correlation functions of the random potentials must be determined in a similar way. This illustrates that the change of variables from the matrix-model data to the ensemble-averaged quantum-mechanical description is highly nontrivial: the disorder distribution is fixed order by order by matching multi-boundary amplitudes.
%\sout{Finally, the functions $W_n$ are uniquely determined by requiring that the perturbative genus expansion in (\ref{genusn}) on the matrix-model side be exactly reproduced by the ensemble-averaged quantum mechanics defined above.} 
The matching provides a stringent consistency condition and ensures the equivalence between the two descriptions at the perturbative level.

%In the Appendix, we  also explain how the perturbative expansion in the framework of ensemble-averaged quantum mechanics reproduces the genus expansion in (\ref{genusn}), and we show how the functions $W_n$ are determined. 
The leading-order contribution $Z_{0,1}$ is obtained from the limit
\be
Z_{0,1}(\beta) =\text{$\lim_{\hbar\rightarrow 0}$}\, \hbar\, {\rm tr} \, e^{-\beta (\hbar^2 %e^{-2 S_0} 
p^2_X +e^{-X/\hbar} + W_0(X))}\,.
\ee
As shown in \cite{Bak:2025eul}, this expression is solved by
\begin{equation} \label{LCPot}
2\pi X = \sqrt{W_0(X)}\, I_{1}\Big(2\pi \sqrt{W_0(X)}\Big)\,,
\end{equation}
which follows from an Abel integral equation for $X\ge 0$. 
(One may further impose $W_n =0$ for $X<0$ at least perturbatively.) Thus $W_0(X)$ is determined uniquely.
In the Appendix, we further demonstrate that $Z_{1,1}(\beta)$ is correctly reproduced within this framework, up to the expected renormalization effects. This agreement is far from trivial: the required input data, namely the correlation function $C_{11}(X,Y)$ and the function $W_1$ are already fixed. %independently. 
This matching therefore provides a nontrivial and additional consistency check of the framework.  Further checks of higher-order contributions $Z_{g,n}$ 
%for $n>1$ 
as well as of the associated topological recursion relations, are required for a %complete 
full validation of the construction. We leave these investigations for future work.
 
To compute physical quantities of interest, such as the complexity, we use the so-called quenched version,
\be
\langle O \rangle_q =\left\langle \frac{{\rm tr} \, O  e^{-\beta H} }{{\rm tr}\, e^{-\beta H}} \right\rangle_{\vec{v}}\,,
\ee
where the quantum-mechanical trace is performed first and the ensemble average is taken afterward. In this quenched average, the random potential in a given realization is fixed in time during the quantum-mechanical evolution. This, for instance, leads to plateau behavior in the presence of the random potential for a given sample. After performing the ensemble average, the plateau behavior remains while fluctuations are averaged out. This quenched version is the one we consider in this work for any physical observables. %In \cite{Bak:2025eul}, 
In \cite{Bak:2025eul}, we checked numerically, for a fixed random-potential sample, that the geodesic length, identified with the Krylov spread complexity, exhibits the ramp--top--slope--plateau behavior\footnote{Some related works include~\cite{Saad:2018bqo,Maxfield:2020ale,Lin:2022rbf,Iliesiu:2021ari, Iliesiu:2024cnh,Boruch:2024kvv,Jafferis:2022wez,Miyaji:2024ity,Bak:2025qgs,Miyaji:2025yvm,Miyaji:2025ucp,Akers:2025ynh}.} as shown in Figure~\ref{fig3}. Here ``top'' and ``slope'' are schematic labels for the late-time finite-entropy crossover of the spread-complexity (wormhole-length) observable, not additional universal semiclassical phases. The slope is controlled by the Heisenberg time, $t_H(E)=2\pi\rho(E)$, and hence $\Delta t_{\mathrm{slope}}=O(t_H)=O(e^{S_0})$, up to non-universal and sample-dependent factors. Since this Krylov spread complexity is distinct from the computational/geometric complexity entering Susskind's opacity criterion \cite{Susskind:2015toa}, its decrease during the slope should be understood as relaxation toward the plateau, rather than as evidence for a robust semiclassical firewall.

In the so-called annealed average, one performs the ensemble average first and then takes the quantum-mechanical trace,
\be
\langle O \rangle_a = \frac{{\rm tr} \, O \langle e^{-\beta H} \rangle_{\vec{v}} }{{\rm tr}\, \langle e^{-\beta H} \rangle_{\vec{v}} } \,.
\ee
Due to the averaging over the potential first, the random potential is no longer fixed in time in the path-integral formulation. Since the realization of the random potential can change over time, the spectrum becomes effectively continuous and full phase cancellation between different states never occurs. Consequently, in this annealed case, the complexity grows linearly in time indefinitely and never reaches a plateau. In this work, we do not use the annealed average for any physical observables.

For the free energy of the system, one is required to compute
\be
\langle  \ln Z_H\rangle_{\vec{v}}=\lim_{n\rightarrow 0}%\frac{1}{n}
\big( \langle   Z^n_H\rangle_{\vec{v}}-1 \big)/n\,,
\ee 
which corresponds to the quenched prescription. The expression on the right-hand side can be evaluated using the replica trick, as is well known in disordered systems. We emphasize that
\be
\langle  \ln Z_H\rangle_{\vec{v}}\neq   \ln \langle Z_H\rangle_{\vec{v}}\,,
\ee 
and therefore care must be taken in this respect.

%%%%%%%%%%%%%%%%%%%%%%%%%
%\section{Introduction}\label{sec0}
\section{Resolution of black hole singularities}\label{sec4}
%%%%%%%%%%%%%%%%%%%%%%%%%
%%%%%%%%%%%%%%%%%%%%%%%%%

If we introduce the extra confining potential $W$, % \cite{Bak:2025eul}, 
the boundary dynamics is significantly modified once this potential becomes relevant, which occurs at time scales
$t \sim {O}( e^{S_0})$ or equivalently $\chi \sim -{O}(e^{S_0})$.  The total potential takes the form
\be
V(\chi)= \frac{e^{2\chi}}{2{\cal C}}  + W\,.
\ee
Accordingly, the equation of motion in~(\ref{orieom}) becomes
\be\label{eom}
\C {\ddot\chi} =\C  {G''} \dot\tau= \C G {G''}= -\frac{e^{2 \chi}}{\C}-\partial_\chi W\,.
\ee
The physical picture is transparent. 
At early times, the effect of the confining potential $W$ is negligible, and the evolution is governed primarily by the Liouville term. 
As $-\chi$ increases, the velocity $\dot{\chi}$ approaches an approximately constant value and remains nearly constant until 
$\chi \sim -{O}(e^{S_0})$.  %This corresponds the well-known linear growth regime of complexity.
This regime corresponds to the well-known linear growth of complexity. 
Once $\chi$ reaches this exponentially large negative value, however, the confining potential becomes important and qualitatively alters the subsequent evolution.

As time evolves further, the system eventually reaches the left turning point $\chi_{ltp}\sim - {O}(e^{S_0})$, at which the velocity vanishes,
$\dot\chi={G'}=0$. %becomes zero. 
At this point,  $\dot\tau  =G=e^{\chi}/\C$ attains its minimum value. Therefore,
\be
\dot\tau \ge \frac{e^{\chi_{ltp}}}{\C}\,. 
\ee
This should be contrasted with the unmodified vacuum branch, where $\dot\tau=\cos\tau$ and the trajectory approaches $\tau=\pi/2$. Once the confining potential is relevant, this vacuum relation no longer applies. The modified dynamics gives $\dot\tau=e^\chi/\C>0$, so the unwrapped global time continues to increase beyond the vacuum endpoint.
From the viewpoint of the cutoff trajectory, this behavior corresponds to an effective resolution of the would-be singularity.
 At the turning point, the spread complexity reaches its maximal value, followed by the slope region. This slope is a Heisenberg-time crossover of the Krylov spread complexity and should not be identified with a separate semiclassical firewall phase.
As we argue below, beyond this point the classical description of the boundary evolution gradually ceases to be valid: 
when the spread complexity of the quantum state becomes of order $e^{S_0}$, quantum effects become significant and the classical approximation begins to break down.
%As will be argued below, after the top region, the classical nature of the boundary evolution begins to break down when the spread complexity of the quantum state becomes of order $e^{S_0}$.

%$\bullet$ phase cancellation plus plateau

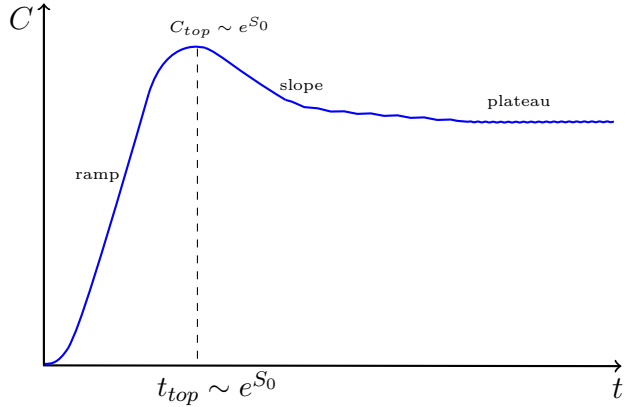
\begin{figure}
\vskip0.3cm 
\begin{center}
\begin{tikzpicture}[x=1cm,y=1cm]

% axes
\draw[thick,->] (-1.01,-1) -- (6.6,-1.01);
\draw[thick,->] (-1,-1) -- (-1,3.8);

\node at (6.55,-1.3) {$t$};
\node at (-1.3,3.6) {$C$};

% --- 1) exponential onset (very short, strong curvature)
\draw[blue,  thick]
  (-0.99,-0.98)
  .. controls (-.902,-0.98) and (-.815,-0.98) ..
  (-0.6792,-0.7678);

% --- 2) fast linear growth
\draw[blue, thick]
  (-0.68,-0.77)
  .. controls (-0.6,-.603) and (-0.5,-.48) ..
  (0.378,2.6);

% --- 3) flat top (almost horizontal)
\draw[blue, thick]
  (0.378,2.6)
  .. controls (0.62,3.38)  and (1.12,3.21)  ..
  (1.12,3.21);

% --- 3) flat top (almost horizontal).
\draw[blue, thick]
  (1.12,3.21)
  .. controls (1.3,3.15) and (1.52,2.92) ..
  (2.18,2.52);

% --- 4) slope  
\draw[blue, thick,decorate, decoration={zigzag, amplitude=0.1mm}]
  (2.18,2.52)
  .. controls (2.62,2.34) and (2.86,2.38) ..
  (4.58,2.23);

%% --- 4) slope  - plateau
%\draw[blue, thick,decorate, decoration={snake,amplitude=0.2mm}]
%  (2.82,2.23)
%  .. controls (3.2,2.21) and (3.3,2.24) ..
%  (3.62,2.23);
  
%  
%% --- 4) slope - plateau
%\draw[blue, thick,decorate, decoration={snake,amplitude=0.12mm,segment length=1.2mm}]
%  (3.62,2.23)
%  .. controls (4.12,2.22) and (4.26,2.23) ..
%  (4.58,2.23);

% --- 5) long plateau
\draw[blue, thick, decorate, decoration={snake,amplitude=0.05mm,segment length=1mm]}]
  (4.58,2.23) .. controls (5.12, 2.22) and (5.26,2.23) ..  (6.5,2.23);

\node at (0.14, 1.5) [left]  {\tiny ramp} ;
\node at (1.32, 3.2) [above]  {\tiny $C_{top} \sim e^{S_{0}}$ } ;
\node at (1.96, 2.7) [right]  {\tiny slope} ;
\node at (5.25, 2.25) [above]  {\tiny plateau} ;

\draw[dashed] (1.02,3.2) -- (1.02,-1);

\node  at (1.29, -1.3) {\small $t_{top}\sim e^{S_{0}} $}; 
\end{tikzpicture}

\end{center}
\vskip-0.3cm 
\caption{Schematic diagram for complexity vs time. The typical behavior of ramp, top, slope and plateau of the complexity  or equivalently the geodesic length is depicted. }\label{fig3}
\end{figure}

%Let us note that with the random potential contribution, the spectrum follows random-matrix statistics, while the level spacing becomes of order $e^{-S_0}$, exhibiting level repulsion. As the time reaches order $e^{S_0}$, the system begins to deviate from purely classical motion, especially beyond the top region. Of course, the so-called Ehrenfest theorem is still obeyed.
In particular, once the random potential is included, the spectrum exhibits random-matrix statistics: the level spacing is of order $e^{-S_0}$ and displays level repulsion. 
Accordingly, at time scales $t \sim O(e^{S_0})$, the evolution begins to deviate from purely classical motion. Nevertheless, the Ehrenfest theorem continues to hold.
We now consider the Hamiltonian 
%$H=\frac{p_q^2}{2\C} + V(q)$ with $V(q)=W+ \frac{e^q}{2\C}$, let us introduce  
\be
H=\frac{p_q^2}{2\mathcal{C}} + V(q),
\qquad 
V(q)= W + \frac{e^{q}}{2\mathcal{C}},
\ee
%the eigenvalues $E_n$ with the corresponding eigenstate $|n\rangle$. 
with eigenvalues $E_n$ and eigenstates $|n\rangle$. 
The thermofield double (TFD) state evolved by $H_{\mathrm{tot}}=2H$ is
%Then the dual TFD state evolved by $H_{tot}=2H$ is given by
\be
|\psi(t)\rangle_{tfd} =\frac{1}{\sqrt{Z}}\sum_n e^{-(\frac{\beta}{2}+2it) E_n}|n,n\rangle \,.
\ee
%with $|n,m\rangle$ denoting $|n\rangle_l \otimes  |m\rangle_r$. 
where $|n,m\rangle \equiv |n\rangle_l \otimes |m\rangle_r$.

%We have 
The expectation values satisfy
\bea
&& \frac{d}{d t} \langle q \rangle_{tfd} =-i \langle [q, 2H] \rangle_{tfd}= \frac{2}{\C} \langle p_q \rangle_{tfd}\,,  \nn
&& \frac{d}{d t} \langle p_q \rangle_{tfd} =-i \langle [p_q, 2H] \rangle_{tfd}= - 2 \langle \partial_{q} V \rangle_{tfd}\,,
\eea 
%leading to  the quantum version of the evolution equation
which together imply the quantum analogue of the classical equation of motion,
\be
\C \frac{d^2}{d t^2} \langle \chi \rangle_{tfd}=- \langle \partial_\chi V \rangle_{tfd}\,.
\ee
More explicitly,
\be
%2 \langle p \rangle_{tfd}=-i \langle [q, H] \rangle_{tfd}
\frac{d}{d t} \langle q \rangle_{tfd} 
= \frac{2i}{Z}\sum_{m,n} e^{-\frac{\beta}{2}(E_m+E_n)} e^{2it(E_m-E_n)} (E_m-E_n)  
\langle m,m| q |n,n \rangle \,.
\ee
%Then for the regime with $t \gg e^{S_0}$, the above becomes zero due to the 
%random erratic phase cancellation, corresponding to %marking 
%the plateau region~\cite{Cotler:2016fpe, Susskind:2018pmk,Saad:2018bqo} with %\cite{Cotler:2016fpe}
For $t \gg e^{S_0}$, the rapidly oscillating phases $e^{2it(E_m-E_n)}$ undergo erratic cancellations due to the random-matrix nature of the spectrum. 
As a result, the off-diagonal contributions ($m\neq n$) average to zero, yielding
\be
\frac{d}{d t} \langle q \rangle_{tfd} \sim 0\,,
\ee
which characterizes the plateau regime~\cite{Cotler:2016fpe,Susskind:2018pmk,Saad:2018bqo}.
%and hence $\frac{d}{d t} \langle q \rangle_{tfd} \sim 0$ in this regime. 
This is also consistent
with the fact $\langle q \rangle_{tfd}$ becomes time independent in the plateau regime
ignoring those erratic phase canceling contributions; this can be explicitly shown as
\be\label{plateau}
\langle q \rangle_{tfd}=\frac{1}{Z}\sum_{m,n} e^{-\frac{\beta}{2}(E_m+E_n)} e^{2it(E_m-E_n)} 
\langle m,m| q |n,n \rangle \sim \frac{1}{Z}\sum_{n} e^{-{\beta}E_n} 
\langle n,n| q |n,n \rangle\,,
\ee
where the last relation follows by considering the phase cancellation of the middle expression
for $m\neq n$. In this regime, one may also show that  $-i \langle [p_q, 2H] \rangle_{tfd}= - 2\langle \partial_q V  \rangle_{tfd}\sim 0$ by a similar argument. 
Thus, the behavior of the complexity near the plateau region admits a natural quantum-mechanical explanation: 
at late times, phase decoherence induced by random-matrix level statistics suppresses coherent evolution, leading to an effective freezing of the expectation values.
%Hence, the behavior of the complexity around the plateau region can be explained.
%As we have shown in the above, $F_-=-\frac{1}{\C}\langle e^q \rangle_{tfd} < 0$ becomes dominant in the beginning. As time goes, the other (positive) contribution $F_+=-2\langle \partial_q W \rangle_{tfd} > 0$ grows. As $F_-$ is getting negligibly smaller while $F_+$ still remains also negligible, $-\langle q \rangle_{tfd}$ grows linearly in time as was shown in the above.  As far as  $-\langle q \rangle_{tfd}$ grows, $F_-$ keeps decreasing while $F_+$ is increasing. As $F_+$ becomes O(1), the growth of $-\langle q \rangle_{tfd}$ begins to decelerate.  It is rather clear that, at some point, %$F_+$ becomes equal to $F_-$ where $-\langle q \rangle_{tfd}$ reaches the maximum. At this maximum point, it is clear that $F_+ > -F_-$ as $F_+$ contribution is winning over $-F_-$ for large $-\langle q \rangle_{tfd}$ as $-F_-$ becomes negligible. Then it turns around, $F_+$ is decreasing while $-F_-$ is growing. Eventually it stops as in the above where $\langle p \rangle_{tfd} \sim 0$ and $F_-+F_+ \sim 0$. (In principle, there could be some oscillations before reaching this point.) 

%If this result is applied to the boundary dynamics, it begins to deviate from the purely classical behavior after the top region. The vacuum expectation value of the geodesic length 
Applying the above result to the boundary dynamics, we find that the evolution begins to deviate from purely classical behavior once the system passes the top region. 
In particular, the vacuum expectation value of the renormalized geodesic length, $\langle \ell_{ren}   \rangle_{tfd}=-2\langle \chi   \rangle_{tfd}$ 
%then starts to decrease and eventually settles down to the plateau value. 
starts to decrease and eventually approaches a constant plateau value.
%Thus, in the plateau regime, the boundary trajectory remains almost parallel to spatial infinity, although it exhibits complicated noise due to randomness. This behavior is depicted in Figure~\ref{fig3}. 
In this late-time regime, the boundary trajectory remains nearly parallel to spatial infinity. 
Although small, irregular fluctuations persist due to the underlying random-matrix dynamics, the overall motion is effectively frozen at the level of expectation values. 
This qualitative behavior is illustrated in Figure~\ref{fig3}.
%\sout{Furthermore, since $\langle \dot\tau   \rangle_{tfd}=\langle e^\chi   \rangle_{tfd}/\C\sim {\rm const.}$, 
%it follows that $\tau(t)$ continues to grow linearly with $t$ even in the plateau regime. Consequently, the boundary global time $\tau(t)$ never stalls, and the would-be singularity is completely avoided.}
Furthermore, since $\langle\dot\tau\rangle_{tfd}=\langle e^\chi\rangle_{tfd}/{\cal C}$ approaches a positive constant in the plateau regime, up to small erratic fluctuations, the unwrapped global time $\tau(t)$ continues to grow approximately linearly with $t$. Consequently, $\tau(t)$ is not bounded in the modified theory, and the would-be singularity is avoided.
The resolution of the singularity is therefore dynamical in nature and is ultimately governed by the behavior of the spread complexity\footnote{In a recent paper~\cite{Balasubramanian:2026azk}, the spread complexity is also used to compute the Hilbert space dimension.}.
%$\tau(t)$ increases linearly in $t$ indefinitely in this regime. Thus, the singularity is fully resolved, and its essence lies in the dynamics of the spread complexity.

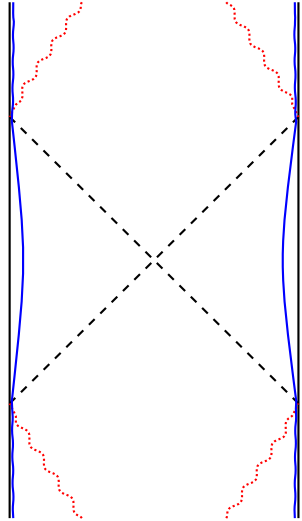
\begin{figure}  
\vskip0.3cm 
\begin{center}
%
%
%
%%%%%%%%%%%%%%%%%%%%%%%%%%%%%%%%%%%%%%%%%%%%%%%%
\begin{tikzpicture}[scale=1.9]
%\draw[thick,blue %,-<
%](-2,0)--++(0,2);
%\draw[thick,blue] (-2,2)%node[left]{\large$t_l$}
%--++(0,2);
%\draw%[thick]
%(-3,0)--++(0,3);
%\draw%[thick]
%(-1.3,0)--++(0,3);
\draw[thick]
(1.7,-0.3)--(1.7,3.3);
\draw[thick]
(3.7,-0.3)--(3.7,3.3);
\draw[thick,dashed] (3.7,0.5) -- (1.7,2.5) ;
\draw[thick,dashed] (1.7,0.5) -- (3.7,2.5) ;

\draw[densely dotted, thick,red,decorate, decoration={snake, amplitude=.4mm}] (2.2,3.3) .. controls (1.8,2.7) .. (1.7,2.5) ;
\draw[densely dotted, thick,red,decorate, decoration={snake,amplitude=.4mm}] (3.2,3.3) .. controls (3.6,2.7) .. (3.7,2.5)   ;
\draw[densely dotted, thick,red,decorate, decoration={snake,amplitude=.4mm}] (2.2,-0.3) .. controls (1.8, 0.3) .. (1.7,0.5) ;
\draw[densely dotted, thick,red,decorate, decoration={snake,amplitude=.4mm}] (3.2,-0.3) .. controls (3.6,0.3) .. (3.7,0.5) ;

\draw[thick,blue,decorate, decoration={snake,amplitude=.05mm}] 
(1.715,2.46) .. controls (1.718,2.5) .. (1.725,3.3);
\draw[thick,blue] 
(1.715,0.515) .. controls (1.82,1.5) .. (1.716,2.465);
\draw[thick,blue,decorate, decoration={snake,amplitude=.05mm}] 
(1.725,-0.3) .. controls (1.718,0.5) .. (1.715,0.52) ;

\draw[thick,blue,decorate, decoration={snake,amplitude=.05mm}] 
(3.685,2.46) .. controls (3.682,2.5) .. (3.675,3.3);
\draw[thick,blue] 
(3.685,0.515) .. controls (3.56,1.5) .. (3.684,2.465);
\draw[thick,blue,decorate, decoration={snake,amplitude=.05mm}] 
(3.675,-0.3) .. controls (3.682,0.5) .. (3.685,0.52);

\end{tikzpicture}
%%%%%%%%%%%%%%%%%%%%%%%%%%%%%%%%%%%%%%%%%%%%%%%%%%%%%%%%%%

\end{center} 
\vskip-0.3cm 
\caption{The boundary cutoff trajectories, depicted by blue lines,  do not touch the AdS boundaries depicted by black lines, and extend to the upper  and lower sides forever. The dashed diagonal lines denote the would-be horizons and the dotted red lines do the would-be singularities. } \label{fig4}
\end{figure}

%$\bullet$ No longer classical.  Left right correlation: One can send a signal but cannot be decoded. horizon disappears but approximate one remains.

%$\bullet$ Figure 2

%%%%%%%%%%%%%%%%%%%%%%%%%
%\section{Introduction}\label{sec0}
\section{Bulk view of the resolution}\label{sec5}
%%%%%%%%%%%%%%%%%%%%%%%%%
%%%%%%%%%%%%%%%%%%%%%%%%%

In this section, we explore the bulk implications of the singularity resolution. 
We begin by discussing the disappearance of the future horizons and the resulting apparent causal connection between the left and right boundaries. 
We then analyze the corresponding modifications of the bulk dilaton profile. (See \cite{Bogojevic:1998ma} for some related issues.)

As the boundary time coordinates $\tau_{r/l}(t)$ grow indefinitely, the bulk spacetime extends across the entire strip. 
As illustrated in Figure~\ref{fig4}, the left and right future horizons disappear completely. 
In contrast to the original eternal black-hole geometry—where the two boundary trajectories are entirely causally disconnected—the modified geometry allows causal contact between the left and right boundaries. Given this causal connection, it is natural to ask whether it is possible to send a physical signal from one side to the other.

%In this section, we describe the bulk implications of the singularity resolution. We first discuss the disappearance of the future horizons and the corresponding apparent causal connection between the left and right boundaries. We then examine the changes in the bulk dilaton field.

%As $\tau_{r/l}(u)$ coordinates of the boundary trajectories grow indefinitely, the bulk spacetime extends over the strip. As shown in Figure \ref{fig4}, the left and right horizons disappear completely. One also finds that the left and right boundary trajectories are now causally connected, whereas in the original black-hole geometry the two trajectories are causally disconnected  completely. As the two boundaries are causally connected, one may  ask whether it is possible to send a signal from one side to the other.

From the view point of the dual TFD state~\cite{Maldacena:2001kr}, it is clear that one cannot send a physical signal from one side to the other since the left and right systems are noninteracting with each other.  To show this explicitly, let us perturb the Hamiltonian of the right-boundary system by
\be
\delta H_r=J_r(t) {\cal O}_r(t)
\ee
where ${\cal O}_r=1\otimes {\cal O}$ acting on the right hand side only. We try to detect this
change from the left side by considering the expectation value of some operator 
\be
{\cal X}_l={\cal X}\otimes 1
\ee
Its induced change is given by
\be
\delta \langle{\cal X}_l(t)   \rangle_{tfd} =i \int^t_{-\infty} dt'  \langle [\delta H_r(t'), {\cal X}_l(t)] \rangle_{tfd} =0 \,,
\ee 
which vanishes identically since $ [{\cal O}_r(t'), {\cal X}_l(t)]=0$. Hence from the left side, the change in the right side cannot be detected. 

From the bulk perspective, once the horizons disappear the geometry appears transparent, suggesting that a signal emitted from one side may propagate to the other without obstruction.
However, a more careful analysis reveals an important limitation. A signal sent from the right boundary at a time
 $t_s \sim O(1)$ reaches the left boundary only at a doubly exponentially
 late time $t_a(t_s)$ as 
 $\dot\tau=e^\chi /\C \sim e^{-O(e^{S_0})}$ %with $a_s\sim O(1)$
  after the turnaround. In principle, this arrival time can be determined from the boundary dynamics. 
  %It also obeys the inequality $\kappa(t_s) > \kappa_h$ where $t_h=\kappa_h e^{S_0}$  $\kappa_h$ is defined by $t_h=\kappa_h\, e^{S_0}$ with $\tau(t_h)=\pi/2$. Hence $t_a(t_s)> t_h$.
%However, a more careful analysis reveals an important limitation. A signal sent from the right boundary at a time
% $t_s \sim O(1)$ reaches the left boundary only at a parametrically 
% late time $t_a=\kappa(t_s) e^{S_0}$ where $\kappa(t_s)\sim O(1)$. In principle, this arrival time can be determined from the boundary dynamics. It also obeys the inequality
% $\kappa(t_s) > \kappa_h$ where %$t_h=\kappa_h e^{S_0}$  $\kappa_h$ is defined by $t_h=\kappa_h\, e^{S_0}$ with $\tau(t_h)=\pi/2$. Hence $t_a(t_s)> t_h$.
At this doubly exponentially large time scale  $t_a$, the spread complexity of the TFD state has already approached its plateau value. The system is then in a regime of extreme complexity and effective randomness. In such a state, any small perturbation becomes indistinguishable from the intrinsic background fluctuations. Moreover, by the time it reaches the opposite boundary, the signal itself has been thoroughly scrambled  and fully delocalized   over the available Hilbert space.
Therefore, despite the apparent geometric transparency, no practically recoverable information can be transmitted in this manner.
%At the time scale, the spread complexity of state $|\psi(t_a)\rangle_{tfd}$ is around its plateau values and the system becomes extremely complicated and random. Hence with such complexity and randomness, a small signal cannot be discriminated from background noise. 

The situation here is reminiscent of the island scenario~\cite{Almheiri:2020cfm,Kim:2020cds,Balasubramanian:2022gmo}, where complexity likewise plays a crucial role. In the semiclassical description, the island degrees of freedom appear to reside behind the horizon. However, in the full quantum description they are encoded in the radiation degrees of freedom.
This apparent tension with causality in the semiclassical picture is resolved, once computational complexity is taken into account. Although the information is, in principle, present in the radiation, extracting it requires decoding operations of order ${\cal O}(e^S)$  complexity. Such operations are exponentially difficult in the entropy and therefore effectively inaccessible in practice.
In this way, complexity restores operational causality: while the information is not fundamentally lost, it cannot be retrieved by any feasible physical process.

%At this time scale, the spread complexity of the TFD state is around its plateau value, and the system becomes extremely complicated and effectively random. With such complexity and randomness, a small signal cannot be distinguished from the background noise. Note also that, around the arribal time, the signal itself becomes scrambled all over the available Hilbert space rather completely. Hence, in practice, no information can be transferred in this manner.  The situation here is reminiscent of the island scenario, in which complexity also plays an essential role. In the semiclassical description, the island degrees of freedom appear to remain inside the horizon, whereas they in fact belong to the radiation degrees of freedom. The apparent clash with causality in the semiclassical description may again be resolved by the fact that decoding these degrees of freedom from the radiation requires operations of order $O(e^S)$ complexity. 

We now turn to the corresponding bulk description of the dilaton field, incorporating the modification of the boundary dynamics. Note that, in the presence of the energy–momentum tensor $T_{ab}$, the dilaton 
satisfies
\be
 \nabla_a \nabla_b \phi -g_{ab} \nabla^2\phi + g_{ab} \, \phi =- T_{ab} \,.
\ee
Assuming that the extra matter sector introduced above is independent of the dilaton field, the bulk spacetime remains AdS$_2$,  with the metric taken in its global form as given in (\ref{global}).
Motivated by the form in (\ref{dilatona}), we assume that the dilaton takes the form
\be
\phi =\C %\bar\phi 
\left(\frac{G(\tau)}{\cos \mu} -\cos \mu \, K(\tau)\right) \,,
\ee
where $G(\tau)$ is given by the solution of (\ref{eom}), whose explicit form can, in principle, be obtained once $W$ is specified. Using the above dilaton equation, one finds
\bea
T_{\tau\tau}/\C &&=-2  K \cos\mu \,, \nonumber\\
T_{\tau\mu}/\C &&=-2 K'  \sin \mu  \,, \nonumber\\
T_{\mu\mu}/\C &&=-\frac{G+ G''-2K}{\cos\mu }
- (K- K'') \cos \mu \,. %\nonumber\\
\eea
%This energy momentum tensor is  covariantly conserved 
Since we do not want the energy-momentum tensor to have a singular contribution along the cutoff trajectories, the $1/\cos \mu$ term in $T_{\mu\mu}$ must be absent. This requirement leads to the condition
\be
K(\tau)=\frac{1}{2} \big(G+%\partial^2_\tau 
G''\big)=-\frac{1}{2} e^{-\chi}\partial_\chi W
\ee
where $\chi$ is related to $\tau$ by the relation $e^\chi =\C G(\tau)$.
Hence, the dilaton and the bulk energy-momentum tensor are uniquely determined once the left confining potential is specified.

A few comments are in order. First, in JT theory without modification, the bulk force is attractive, and any mass inevitably falls into the horizon. As noted in \cite{Susskind:2019ddc}, this attractive nature is related to the linear growth of the spread complexity in the ramp region. Without our modification, the linear growth continues, as we have already mentioned. The relevant force in this case is given by
\be
F_- = -\frac{\langle e^{2\chi} \rangle}{\C}\,,
\ee
which is negative definite. The turnaround is controlled by a new force,
\be
F_+ = -\langle \partial_\chi W \rangle \,,
\ee
which is positive definite (in an averaged sense over the random contributions). Hence, in this regime, around the top, the dilaton becomes negative in the bulk, while remaining positive at the boundary. 

In the plateau region, the expectation values of any operator become constant, using the same argument as in (\ref{plateau}). Hence, in this region,
\be
\phi = \C \left(\frac{a}{\cos \mu} -\cos \mu \, b \right)\,,
\ee
where $a=\langle G \rangle_{plateau}$ and  $b=\langle K \rangle_{plateau}$. Therefore, in the plateau regime, the dilaton becomes stationary in the 
$\tau$ coordinate. A precise estimation of 
$a$ and $b$ as functions of time $t$ would be of interest, but we leave this for future work.

Next, we turn to the discussion of the bulk energy. This can be constructed from $T_{ab}$ using covariant conservation:
\be
\nabla_a (T^a\negthinspace_b f^b)=0\,,
\ee
where $f= \partial_\tau$ is the Killing vector satisfying the Killing equation
$\nabla_a f_b+\nabla_b f_a =0$. This implies the ordinary conservation law
\be
\partial_a (\sqrt{-g} T^a\negthinspace_\tau)=0\,,
\ee
from which one can define the bulk energy as
\be
E_{bulk}=-\int^{\pi/2}_{-\pi/2} d\mu \sqrt{-g} \langle T^\tau\negthinspace_\tau \rangle=-4 \C \langle  K \rangle \,,
\ee
which is negative definite. Hence, the bulk energy induced by the left confining potential is negative. This negative-energy contribution becomes relevant at times
 $t\sim O(e^{S_0})$ resulting in %deriving 
 the singularity resolution and the turnaround.

Finally, using the energy–momentum tensor, one may verify that
\be
-{\cal C}\, \frac{d}{dt} {\rm Sch}\left(\tan \frac{\tau(t)}{2}, t\right) =(\dot\tau)^2 T_{\tau \mu} 
\ee
holds along the cut-off boundary. This relation may be interpreted as the bulk force, induced by the modification of the energy–momentum tensor, acting on the boundary trajectories. It therefore leads to a corresponding modification of the boundary dynamics.

%%%%%%%%%%%%%%%%%%%%%%%%%
%\section{Introduction}\label{sec0}
\section{Conclusions}\label{sec6}
%%%%%%%%%%%%%%%%%%%%%%%%%
%%%%%%%%%%%%%%%%%%%%%%%%%

%Not purely classical. Higher D generalization.

In this work, we have demonstrated that introducing a left confining 
potential in JT gravity provides a natural and dynamical mechanism for 
resolving black hole singularities. Requiring a discrete 
bulk spectrum with random statistics naturally leads to the existence of 
a left confining potential which becomes relevant only when the renormalized 
geodesic length is of order $e^{S_0}$. 
In the absence of this potential, the boundary cutoff trajectory asymptotes to
the AdS boundaries, corresponding to an ever-growing wormhole length and the 
formation of a bulk singularity. With the confining term included, however, 
the evolution develops a turning point at $\chi \sim -O(e^{S_0})$, 
where the effective repulsive force prevents the indefinite growth of the 
wormhole length. Then, the boundary trajectory continues to evolve 
beyond the would-be endpoint, eliminating the geometric pathology.
The singularity resolution is therefore not imposed by hand but emerges 
dynamically from the discreteness of the spectrum. The Krylov spread 
complexity, which is identified with the wormhole length, then settles
into a plateau after reaching a maximum, rather than growing indefinitely.

From the bulk perspective, this modification drastically alters the causal 
structure of spacetime. The would-be future horizons disappear and the left
and right boundaries become causally connected. However, we argued that,
in practice, no information can be transferred between them.
Any signal sent from one boundary reaches the other only at doubly exponentially late time scales, 
%$t \sim O(e^{S_0})$, 
a regime where the system becomes extremely complicated
and effectively random. At this stage, the signal is thoroughly scrambled 
over the Hilbert space, making it indistinguishable from the background. 
Thus, the semiclassical paradox of causality violation is resolved by the 
extreme complexity of the quantum state. The ``complexity barrier'' then
emerges as the quantum successor to the classical event horizon. 

Our results suggest that the resolution of the singularity is an inherently 
non-classical phenomenon, arising directly from the discrete and random
nature of the underlying spectrum. The smooth geometric description breaks down 
at time scale of $O(e^{S_0})$ in the plateau regime, where the physics is 
inaccessible in semiclassical analyses and is governed by the random 
statistics of the energy levels. 

It is useful to compare our late-time mechanism with the baby-universe tunneling process of  Stanford and Yang \cite{Stanford:2022fdt}. Their genus-one handle-disk contribution becomes important in the same parametric regime, $t\sim e^{S_0}$, in which the confining potential affects the length dynamics. We %therefore 
do not view the confining potential as eliminating this baby-universe channel. Rather, the genus-one calculation describes a particular topology-changing Lorentzian process, while the confining potential gives the effective finite-entropy completion that regulates the unbounded growth of the wormhole length and drives the spread complexity toward the plateau. In the multi-boundary quantum-mechanical formulation, such topology-changing effects are encoded in the connected amplitudes $Z_{g,n}$, or equivalently in the cumulants of the random confining potential. A direct Lorentzian computation of the corresponding transition between expanding and contracting components in our quantum-mechanical formulation 
would be an interesting future direction.

Several important questions remain for future study. 
A full verification of the framework requires matching higher-order 
contributions $Z_{g,n}$ and establishing compatibility with topological 
recursion beyond the leading orders.
The bulk energy associated with the left confining potential turns
out to be negative, and becomes relevant at times $t \sim O(e^{S_0})$. 
It would also be interesting to clarify the precise bulk interpretation of 
the random potential. While this study focused on 
two-dimensional gravity, it would be of great interest to investigate 
whether similar mechanisms can be generalized to higher dimensions.

\subsection*{Acknowledgement}
% We would like to thank Jong-Hyun Baek, Jeongwon Ho, and O-Kab Kwon for enlight
%ening discussions. 
% D.B. was supported by the 2024 Research Fund of the University of Seoul. 
DB was supported in part %by NRF Grant RS-2023-00208011, and
 by Basic Science Research Program through NRF funded by the Ministry of Education (2018R1A6A1A06024977). 
C.K. was supported by NRF Grant 2022R1F1A1074051.
 S.-H.Y. was supported in part by NRF grant funded by the Korea government(MSIT) RS-2025-00553127.

\appendix
{\center \section*{Appendix}}
%\hskip1cm

\section{Perturbative expansions of the partition function} \label{AppA}
%%%%%%%%%%%%%%%%%%%%%%%%%
 %%%%%%%%%%%%%%%
\renewcommand{\theequation}{\thesection.\arabic{equation}}
  \setcounter{equation}{0}
We consider a one-dimensional quantum system with Hamiltonian $H=p^2+V(X)$ and aim to compute the canonical partition function
\be 
Z_H = \int dX \, \langle X| e^{-\beta H}|X\rangle
\ee
as an expansion in powers of $\hbar$. To this end, we introduce the Wigner transform of an operator $\hat O$,
\begin{equation}
O_W(X,p)
:= \int_{-\infty}^{\infty} dy\;
e^{-\frac{i}{\hbar}py}\,
\left\langle X+\frac{y}{2}\Big|\hat O\Big|X-\frac{y}{2}\right\rangle .
\end{equation}
For $\hat O = e^{-\beta H}$, the partition function can be written in phase-space form as
\be
Z_H = \frac{1}{2\pi \hbar}\int dX dp\, (e^{-\beta H})_W (X,p)\,.
\ee
It is useful to examine the symmetry properties of the Wigner kernel. Let us define
\begin{align}
K(\hbar, X)
&:= \int dp  dy\, 
e^{-\frac{i}{\hbar} p y}\,
\left\langle X+\frac{y}{2}\Big| e^{-\beta H}\Big|X-\frac{y}{2}\right\rangle .
\end{align}
Under the transformation $p \to -p$, one finds
\begin{align}
K(\hbar, X)
&= \int dp  dy \,
e^{+\frac{i}{\hbar} p y}\,
\left\langle X+\frac{y}{2}\Big| e^{-\beta H}\Big|X-\frac{y}{2}\right\rangle % \nonumber\
%&
= K(-\hbar, X)\,.
\end{align}
This symmetry implies that $K(\hbar,X)$ admits an expansion containing only even powers of $\hbar$. Indeed one may show that the partition function takes the standard Wigner–Kirkwood form~\cite{Hillery:1983ms,Brack:1997}
\begin{equation}
Z_H
= \int \frac{dX dp}{2\pi\hbar}\,
e^{-\beta \big(p^2 + V(X)\big)}
\left[
1 + \sum_{g=1}^{\infty} \hbar^{2g} A_{2g}(X)
\right],
\label{WKgeneral}
\end{equation}
where the coefficient functions $A_{2g}(X)$ are local expressions constructed from $V(X)$ and its derivatives. This expansion provides a systematic semiclassical approximation to the quantum partition function. At order $\hbar^2$, the coefficient is given by the well-known expression
\be
A_2(X)=\frac{1}{12}\left(
\beta^3(V')^2-2 \beta^2 V''
\right)\,. 
\ee
Performing the Gaussian integral over $p$, one finally obtains
\be
Z_H=\frac{1}{\hbar \sqrt{4\pi \beta}}\int dX e^{-\beta V(X)}\left[
1+\frac{1}{12}\left(
 \beta^3(V')^2-2\beta^2 V''
\right) \hbar^2 +O(\hbar^4)
 \right]\,.
\ee
Some explicit expressions for $A_{2g}$ can be further found in \cite{Jizba:2014rvg}.

For the problem of interest, we identify the semiclassical expansion parameter as $\hbar=e^{-S_0}$ and  consider the Hamiltonian
$H=p^2+V(X)$
where the potential has the form $V(X)=W(X)+e^{-X/\hbar}$ and $p$ has the standard form $p=-i \hbar\partial_X$.

The term $e^{-X/\hbar}$ is intrinsically nonperturbative in $\hbar$. Indeed, it does not admit a Taylor expansion around $\hbar=0$, since
\be
\big(\partial/\partial\hbar\big)^n e^{-X/\hbar} |_{\hbar=0}
\ee 
is ill-defined for all $n$. In the semiclassical limit $\hbar\rightarrow 0$, this term generates an infinitely steep potential barrier, effectively forbidding access to the region $X<0$. As a consequence, the nonperturbative contribution $e^{-X/\hbar}$ enforces a boundary condition at $X=0$, while leaving the dynamics in the region $X>0$
unaffected at any finite order in perturbation theory. 

If one is interested purely in the perturbative expansion in powers of $\hbar$,
it is therefore consistent to discard such nonperturbative effects and restrict the configuration space to the half-line $X\in [0,\infty)$. In this approximation, the partition function reduces to
\be
Z_H\cong \frac{1}{\hbar \sqrt{4\pi \beta}}\int^\infty_{0} dX e^{-\beta W(X)}\left[
1+\frac{1}{12}\left(
 \beta^3(W')^2-2\beta^2 W''
\right) \hbar^2 +O(\hbar^4)
 \right] \,.
\ee
%whose full perturbative expansion can be obtained from (\ref{WKgeneral}) by replacing $V$ by $W$ and restricting the integration domain by $[0, \infty)$. 
More generally, the complete perturbative Wigner-Kirkwood expansion follows directly from \eqref{WKgeneral} upon replacing $V(X)$ by $W(X)$ and restricting the integration domain to the half-line $[0,\infty)$. All effects associated with the exponential term $e^{-X/\hbar}$ are nonperturbative in $\hbar$ and are thus consistently omitted within this purely perturbative treatment.

With the perturbative expression for the partition function at hand, we may now perform the ensemble average. The result takes the form
\be 
\langle Z_H(\beta)\rangle_{\vec{v}} \cong Z_{0,1}\, \hbar^{-1} +\sum^\infty_{g=1} Z_{g,1}\hbar^{2g-1}
\ee
It is manifest that the ensemble average generates only additional even powers of $\hbar$, so that the final result naturally organizes itself into a genus expansion.

At leading order, one finds
\be
Z_{0,1}=\frac{1}{\sqrt{16\pi\beta^3}} e^{\frac{\pi^2}{\beta}}=\frac{1}{\sqrt{4\pi \beta}} \int_0^\infty dX e^{-\beta W_0(X)}\,.
\ee
This expression can be inverted by an inverse Laplace transform, yielding the potential $W_0(X)$ given in \eqref{LCPot}. A detailed derivation of this inversion was presented in the appendix of Ref.~\cite{Bak:2025eul}.

One may also compute the connected two-boundary partition function $Z_{0,2}(\beta_1, \beta_2)$ directly on the quantum-mechanical side. It is given by
\be
Z_{0,2}(\beta_1, \beta_2)=\frac{\sqrt{\beta_1 \beta_2}}{4\pi }\int^\infty_{0} \negthinspace\negthinspace\negthinspace dX\,W_1(X) \int^\infty_{0}  \negthinspace\negthinspace\negthinspace dY\,W_1(Y) e^{-\beta_1 W_0(X)-\beta_2 W_0(Y)} \,C_{11}(X,Y)\,.
\ee
With the known form  $Z_{0,2}=\frac{\sqrt{\beta_1\beta_2}}{2\pi (\beta_1+\beta_2)}$ \cite{Saad:2019lba}, and taking double inverse Laplace transforms of $4\pi \,Z_{0,2}/\sqrt{\beta_1\beta_2}$
 to the variables $W_0(X)$ and $W_0(Y)$ of the both sides, one finds
\be
2\delta \big(W_0(X)-W_0(Y)\big)=\frac{W_1(X)}{W'_0(X)} \frac{W_1(Y)}{W'_0(Y)} \,C_{11}(X,Y) \,.
\ee 
With the choice of $W_1=\sqrt{2 W'_0}$, this equation is solved by the correlator given in (\ref{cw1}).
%\sout{With the translation invariance of the correlation function $C(X)$, this equation admits a unique solution, given by (\ref{cw1}), up to an overall normalization. We fix this normalization to unity.}

At the next order in $\hbar$ in the one-boundary partition function, the coefficient $Z_{1,1}$ is given by
\be\label{z11}
Z_{1,1}(\beta)\negthinspace=\negthinspace\frac{\beta^{\frac{3}{2}}}{4\sqrt{\pi }} \int_0^\infty\negthinspace\negthinspace dX e^{-\beta W_0(X)}\negthinspace \left[ W_1(X)  W_1(X\negthinspace+\negthinspace\alpha \epsilon) \delta_\epsilon (0) +\frac{1}{6}\left(
 \beta (W'_0)^2\negthinspace-\negthinspace 2 W''_0
\right) 
\right]\,.
\ee
Here $\delta_\epsilon(x)=\frac{1}{\epsilon}e^{-\frac{\pi x^2}{\epsilon^2}}$ is a regulated delta function, and $\alpha\epsilon$ denotes the asymmetric point-splitting parameter.
We would like to match this expression with the known result \cite{Saad:2019lba}
\be
Z_{1,1}(\beta)=\frac{\,\,\,\sqrt{\beta}}{12\sqrt{\pi}}(\beta +\pi^2)\,.
\ee
Using the relation $W_1=\sqrt{2 W'_0}$, the right-hand side of  (\ref{z11}) can be rearranged into
\be
\frac{\beta^{3/2}}{4\sqrt{\pi }} \negthinspace \left[ 2\frac{\delta_\epsilon (0) }{\beta}+\frac{1}{3} 
+\big(\alpha\epsilon \delta_\epsilon (0) -1/6 \big)\int_0^\infty\negthinspace\negthinspace dX W''_0 e^{-\beta W_0(X)}\right]\,.
\ee
Reproducing the known answer requires the elimination of the last term in parentheses. This fixes the asymmetric point-splitting parameter through $\alpha\epsilon \delta_\epsilon (0) =1/6$,
which implies  $\alpha=1/6$.
Finally, by subtracting an appropriate local counterterm, the divergent quantity $ \delta_\epsilon (0) $  may be replaced by its renormalized value $r_1=\pi^2/6$. With this prescription, the expression above precisely reproduces the known result for $Z_{1,1}(\beta)$.

To verify the three-point function in \eqref{threep}, we begin with the connected three-boundary amplitude on the matrix-model side, $Z_{0,3}=\sqrt{{\beta_1\beta_2\beta_3}}/{\pi^{3/2}}$.  In the ensemble-averaged quantum-mechanical description, the connected three-boundary contribution consists of two distinct types: one built from two two-point correlators, and the other from the three-point correlator of $v_1$. A straightforward evaluation of these two types reproduces the required  $Z_{0,3}$, thereby verifying the three-point function in \eqref{threep}. This construction can be generalized to genus zero higher connected amplitudes 
$Z_{0,n}$,  which is a topic for future work.
%\begin{align}
%\hbar\, Z_{0,3}
%&=
%-\frac{\sqrt{\beta_1 \beta_2 \beta_3}}{8\pi^{3/2}}
%\int_0^\infty dX\, W_1(X)
%\int_0^\infty dY\, W_1(Y)
%\int_0^\infty dZ\, W_1(Z)
%\nonumber \\
%&\quad \times
%\langle v_1(X) v_1(Y) v_1(Z)\rangle_{\vec v}\, e^{-\big(\beta_1 W_0(X)+\beta_2 W_0(Y)+%%\beta_3 W_0(Z)\big)}.
%\end{align}
%\sout{We require that this result be reproduced by the connected three-boundary contribution computed in the ensemble-averaged quantum-mechanical description.}
%\sout{Substituting the expression \eqref{threep} for the three-point correlator together with$W_1=\sqrt{2W'_0}$,  and taking the limit $\epsilon\rightarrow 0$ at the end of the computation, one verifies that the right-hand side precisely reproduces the matrix-model result.}

One may hope that this procedure can be iterated systematically to higher orders.
% in $g$ and $n$. 
%\sout{In particular, at each order in $\hbar$, the known coefficient $Z_{g,1}(\beta)$ should determine the corresponding correction $w_g(X)$ to the effective potential. 
%If this structure persists,}
The full perturbative expansion of the effective potential could be reconstructed recursively from the genus expansion of the ensemble-averaged partition function. Establishing this recursive relation in a precise manner, however, requires additional consistency checks, which we leave for future investigation.

%Hopefully this procedure might be iterated systematically to higher orders: For instance, at each order in $\hbar$, the known coefficient $Z_{g,1}(\beta)$ determines the next correction $W_g(X)$. In this way, the full perturbative expansion of the effective potential might be consistently reconstructed recursively from the genus expansion of the ensemble-averaged partition function. Further consistency checks are necessary in this direction.

\end{document}